\documentclass[10pt,twocolumn,letterpaper]{article}

\usepackage{iccv}              

\usepackage[T1]{fontenc}
\usepackage{multirow}
\usepackage{subcaption}
\usepackage[dvipsnames]{xcolor}
\usepackage{caption}
\usepackage{flushend}
\usepackage{siunitx}
\usepackage{tipa}
\usepackage{amsmath}
\usepackage{amssymb}

\usepackage{glossaries}
\newacronym{ai}{AI}{Artificial Intelligence}
\newacronym{dl}{DL}{Deep Learning}
\newacronym{asv}{ASV}{Automatic Speaker Verification}
\newacronym{asr}{ASR}{Automatic Speech Recognition}
\newacronym{tts}{TTS}{Text-to-Speech}
\newacronym{vc}{VC}{Voice Conversion}
\newacronym{poi}{POI}{Person-of-Interest}
\newacronym{eer}{EER}{Equal Error Rate}
\newacronym{auc}{AUC}{Area Under the Curve}
\newacronym{snr}{SNR}{Signal-to-Noise Ratio}
\newacronym{ipa}{IPA}{International Phonetic Alphabet}
\newacronym{rfcc}{RFCC}{Rectangular Filter Cepstral Coefficient}
\newacronym{gmm}{GMM}{Gaussian Mixture Model}
\newacronym{moe}{MoE}{Mixture of Experts}
\newacronym{cqcc}{CQCC}{Constant-Q Cepstral Coefficient}
\newacronym{dtw}{DTW}{Dynamic Time Warping}
\newacronym{llm}{LLM}{Large Language Model}

\definecolor{iccvblue}{rgb}{0.21,0.49,0.74}
\usepackage[pagebackref,breaklinks,colorlinks,allcolors=iccvblue]{hyperref}

\def\paperID{16} 
\def\confName{ICCV}
\def\confYear{2025}

\title{Phoneme-Level Analysis for Person-of-Interest Speech Deepfake Detection}

\author{Davide Salvi, Viola Negroni, Sara Mandelli, Paolo Bestagini, Stefano Tubaro\\
Dipartimento di Elettronica, Informazione e Bioingegneria (DEIB), Politecnico di Milano, Italy\\
{\tt\small \{davide.salvi, viola.negroni, sara.mandelli, paolo.bestagini, stefano.tubaro\}@polimi.it}
}

\begin{document}
\maketitle
\begin{abstract}
Recent advances in generative AI have made the creation of speech deepfakes widely accessible, posing serious challenges to digital trust.
To counter this, various speech deepfake detection strategies have been proposed, including Person-of-Interest (POI) approaches, which focus on identifying impersonations of specific individuals by modeling and analyzing their unique vocal traits.
Despite their excellent performance, the existing methods offer limited granularity and lack interpretability.
In this work, we propose a POI-based speech deepfake detection method that operates at the phoneme level.
Our approach decomposes reference audio into phonemes to construct a detailed speaker profile. In inference, phonemes from a test sample are individually compared against this profile, enabling fine-grained detection of synthetic artifacts.
The proposed method achieves comparable accuracy to traditional approaches while offering superior robustness and interpretability, key aspects in multimedia forensics.
By focusing on phoneme analysis, this work explores a novel direction for explainable, speaker-centric deepfake detection.
\end{abstract}    
\section{Introduction}

Recent advancements in generative AI have made it possible to produce hyper-realistic synthetic content with unprecedented ease.
While these technologies hold enormous potential for applications in entertainment, education, and accessibility, they also pose significant risks to security, privacy, and trust in digital communication~\cite{golda2024privacy}.
One of the most concerning menaces is represented by deepfakes, synthetic multimedia content generated through \gls{dl} techniques that depict individuals performing actions or making statements they never actually did~\cite{verdoliva2020media, masood2023deepfakes}.
In the audio domain, speech deepfakes allow the generation of synthetic speech that mimics the voice of a target speaker with remarkable realism, enabling malicious applications such as impersonation, fraud, and disinformation.

To combat the growing risks associated with the misuse of synthetic speech, the multimedia forensics community has developed a wide range of countermeasures, primarily aimed at protecting \gls{asv} systems and enhancing deepfake detection by reliably distinguishing between real and fake speech signals~\cite{rana2022deepfake, cuccovillo2022open}. 
To this end, a variety of approaches have been explored, including transformer-based architectures~\cite{zaman2024hybrid, bartusiak2023transformer, cuccovillo2024audio}, \gls{moe} models~\cite{negroni2025leveraging, wang2024mixture}, and others \gls{dl}-based techniques~\cite{attorresi2022prosody, tak2022automatic, jung2022aasist, yang2024robust, zhang2024audio}.

\begin{figure}[t]
\centering
\includegraphics[width=\columnwidth]{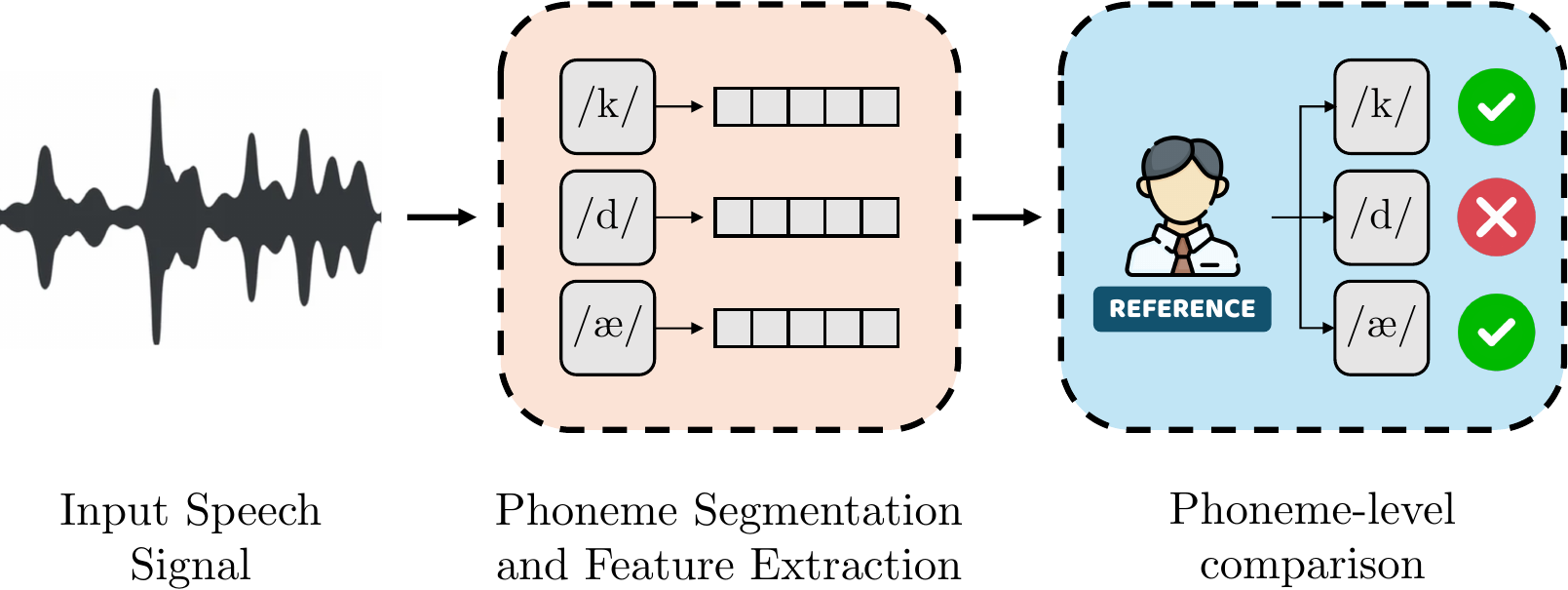}
\caption{Proposed phoneme-based \gls{poi} speech deepfake detection pipeline.}
\label{fig:teaser}
\end{figure}

More recently, \gls{poi}-oriented methods have emerged as a promising strategy to enhance speaker-specific protection~\cite{agarwal2019protecting, cozzolino2023audio}.
These approaches create a speaker profile based on a set of reference recordings from a target individual requiring protection.
During inference, the profile is used to compute the similarity between a given speech sample and the reference speaker's characteristics, determining the authenticity of the analyzed track~\cite{pianese2022deepfake}.
Compared to conventional deepfake detection systems, \gls{poi}-based methods offer two key advantages.
First, their speaker-centric nature leads to higher accuracy in detecting attacks against specific individuals.
Second, since they rely exclusively on real speech data rather than training on synthetic samples, they proved remarkable generalization capabilities and enhanced robustness against various deepfake generation techniques~\cite{zhang2021one}.

One of the most widely used approaches in \gls{poi}-based methods involves representing each speech sample with a single feature vector, obtained using either supervised learning models or pre-trained foundation models~\cite{pianese2024training}.
While this approach has demonstrated strong detection performance, it is inherently limited in granularity.
Representing an entire speech segment with a fixed-length embedding may result in the loss of fine-grained cues that are critical for detecting synthetic speech.
Additionally, this coarse representation reduces interpretability, as it does not provide information about which specific aspects of the speech signal contribute to the final decision. 
As a result, it may be more susceptible to relying on spurious artifacts or shortcuts rather than genuine indicators of synthetic speech~\cite{wang2025asvspoof}.

In this work, we propose a phoneme-level approach to \gls{poi}-based speech deepfake detection, shifting from full-signal analysis to a novel fine-grained phoneme-centric evaluation, as illustrated in Figure~\ref{fig:teaser}.
Rather than processing the entire speech signal as a whole and extracting a single, global feature representation, our method decomposes the input audio into individual phonemes and then processes each phonetic unit independently to extract a dedicated feature vector from it.
These phoneme-level embeddings, derived from reference speech, are aggregated to construct a speaker-specific phoneme profile.
During inference, the test audio is similarly decomposed into phonemes, and each of these is individually compared to the speaker profile, enabling the detection of synthetic artifacts at a much finer temporal and linguistic resolution. 

Our method is grounded on the idea that speech deepfake generators, despite their sophistication, struggle to replicate short, fundamental units of speech, such as phonemes, with perfect fidelity.
By refining the analysis and focusing only on the most critical parts of the signal, our method identifies the specific phonemes that deviate significantly from the reference speaker’s characteristics, resulting in a more robust and reliable analysis.
Moreover, the proposed approach inherits the advantages of one-class \gls{poi}-based methods, making it more resilient to different deepfake generators and diverse recording conditions.

This work explores phoneme-level analysis as a promising approach for \gls{poi}-based speech deepfake detection. 
Our findings support the viability of this method and suggest future research directions, such as integrating phonetic knowledge into end-to-end detectors and developing more interpretable, speaker-centric solutions.
Understanding which specific aspects of speech can be reliably used to discriminate between real and fake audio, and localizing generation artifacts within a signal in both time and frequency domains, could serve as valuable tools to combat voice spoofing in an era of increasingly realistic synthetic media.

The rest of the paper is organized as follows.
Section~\ref{sec:background} provides background on \gls{poi} and phoneme-based speech deepfake detection methods.
Section~\ref{sec:method} formally defines the problem at hand and details the proposed detection pipeline.
Section~\ref{sec:setup} outlines the experimental setup used to evaluate the method.
Section~\ref{sec:results} presents the results along with analysis and discussion.
Finally, Section~\ref{sec:conclusion} summarizes the key findings and outlines directions for future work.
\glsreset{poi}

\section{Background}
\label{sec:background}

\vspace{.5em}\textbf{POI-Based Speech Deepfake Detection.}
The rapid advancement of synthetic speech generation techniques has led to growing interest in developing reliable methods for detecting speech deepfakes.
In recent years, the research community has proposed a wide range of techniques based on diverse detection paradigms~\cite{amerini2025deepfake, li2025survey}, framing the task as a binary classification problem and relying heavily on supervised learning.

While these supervised methods have shown promising results, they also have important limitations.
For instance, accurately identifying speech generated by methods not seen during training remains a significant challenge and undermines the robustness of these systems in real-world conditions.
To address this limitation, \gls{poi}-based methods have emerged as an alternative. These approaches focus on speaker-specific analysis, training exclusively on real data from a given target speaker.
This approach leads to improved accuracy and robustness for the enrolled speaker, as well as enhanced generalization across a wide range of speech generation techniques.

While several \gls{poi}-based deepfake detection methods have been proposed in the video domain~\cite{agarwal2019protecting, agarwal2020detecting, cozzolino2021id, cozzolino2023audio}, their application to audio remains relatively underexplored.
Notable contributions in this space include the work by Pianese et al.~\cite{pianese2022deepfake, pianese2024training}, who introduced a \gls{poi}-based framework for speech deepfake detection that reframes the problem as a speaker verification task: assessing whether the voice in a test sample matches the identity it claims.
This approach falls between the speaker verification and speech deepfake detection tasks, providing strong performance in both fields.
The contamination between \gls{asv} and speech deepfake detection has also been highlighted in other recent studies~\cite{ge2023can, jung24d_interspeech}, which observed that \gls{asv} systems often exhibit an inherent capacity to reject spoofed audio without explicit training for that purpose.

A related and more recent \gls{poi}-inspired approach is presented in~\cite{coletta2025anomaly}, where the authors frame speech deepfake detection as an anomaly detection problem. By training solely on real data, the system learns to detect deviations characteristic of synthetic speech. The use of speaker-specific models in this context closely aligns with \gls{poi} settings and provides compelling evidence for the efficacy of one-class learning in voice spoofing detection.

\vspace{.5em}\textbf{Phoneme-Based Speech Deepfake Detection.}
Phoneme-based speech deepfake detectors are a class of forensic classifiers that leverage the linguistic structure of speech by extracting information tied to phonetic content, which can be used to improve the detection capabilities of the model.

One of the earliest studies to investigate this aspect is~\cite{suthokumar2019phoneme}.
There, the authors analyzed phonetic variations in replay attacks, motivated by earlier works in emotion recognition~\cite{sethu2008phonetic} and speaker verification~\cite{ma2018speaker}, where incorporating phonetic information proved effective.
The underlying idea behind this work is that different phonemes are affected differently by the channel distortions introduced during replay attacks, due to their distinct spectral characteristics.
Specifically, some phonemes emphasize high-frequency energy, while others concentrate on lower frequency bands.
To validate their hypothesis, the authors use \glspl{rfcc} and a phoneme recognizer~\cite{schwarz2006hierarchical} to train phoneme-specific pairs of \glspl{gmm}.
Their analysis identified fricatives, nasals, and silences as particularly informative for detecting replay attacks, aligning with prior findings that showed that replay attacks introduce artifacts primarily in high-frequency regions \cite{nagarsheth2017replay}.

Building on this idea, Dhamyal et al.~\cite{dhamyal2021using} proposed the first explainable \gls{dl}-based approach to study phonetic differences between real and fake speech. 
Inspired by speech recognition techniques~\cite{chan2016listen, palaskar2018acoustic}, they introduced a self-attention mechanism within a SENet model trained on \glspl{cqcc}.
Attention weights were then aligned with annotated phoneme boundaries using \gls{dtw}, enabling frame-level interpretability.
Their findings reinforced the importance of fricatives and nasals but also showed that focusing solely on vowels yielded detection performance that surpassed all individual phoneme classes.

More recently, Zhang et al.~\cite{zhang2025phoneme} proposed a phoneme-sequence-based approach to deepfake detection.
Unlike prior methods that analyze phonemes in isolation, their technique models the temporal sequence of phonemes across an entire utterance.
They fine-tuned a pre-trained \gls{llm} for phoneme recognition and aligned frame-level acoustic features, extracted using a shared encoder, with corresponding phonemes. 
These sequences were then passed to a binary classifier to determine whether the utterance was real or fake.
While this method sacrifices the fine-grained analysis of specific phoneme types, it achieves performance comparable or superior to many state-of-the-art detectors.

In contrast to these prior works, our method leverages phonemes as foundational units in a \gls{poi}-based detection framework.
It is based on the hypothesis that synthetic speech generators struggle to accurately replicate a speaker’s unique phonetic patterns. In practical scenarios, when a reference speaker can provide a large number of real tracks to protect their identity, our method constructs a personalized phoneme dictionary using compact embeddings derived from these genuine samples.
Test recordings can then be analyzed by comparing their phonemes against this dictionary to assess whether they match the target speaker’s genuine speech characteristics or not.
\section{Proposed Method}
\label{sec:method}

In this section, we formalize the \gls{poi} speech deepfake detection task and introduce our phoneme-level detection framework.

\subsection{Problem Formulation}
\label{subsec:problem}

The \gls{poi}-based speech deepfake detection problem is formally defined as follows.
Let $\mathfrak{R}=\{ \mathbf{x}_0, \mathbf{x}_1, \ldots, \mathbf{x}_{I-1}\}$ denote a set of $I$ speech tracks $\mathbf{x}$, all uttered by a target speaker $S$.
These tracks are taken as references to construct a speaker profile, which captures the unique vocal characteristics of $S$.
During inference, we are given a test speech signal $\mathbf{x}$, sampled at a frequency $f_\text{s}$ and associated with a binary class label $y \in \{0, 1\}$, where $y=0$ denotes that $\mathbf{x}$ is an authentic recording of $S$, while $y=1$ indicates that $\mathbf{x}$ is a deepfake attempting to imitate $S$.
The goal of the task is to predict the class label $y$ of the test signal $\mathbf{x}$ by measuring its distance from the speaker profile.
 
\subsection{Proposed System}
\label{subsec:system}

\begin{figure}
    \centering
    \includegraphics[width=0.95\columnwidth]{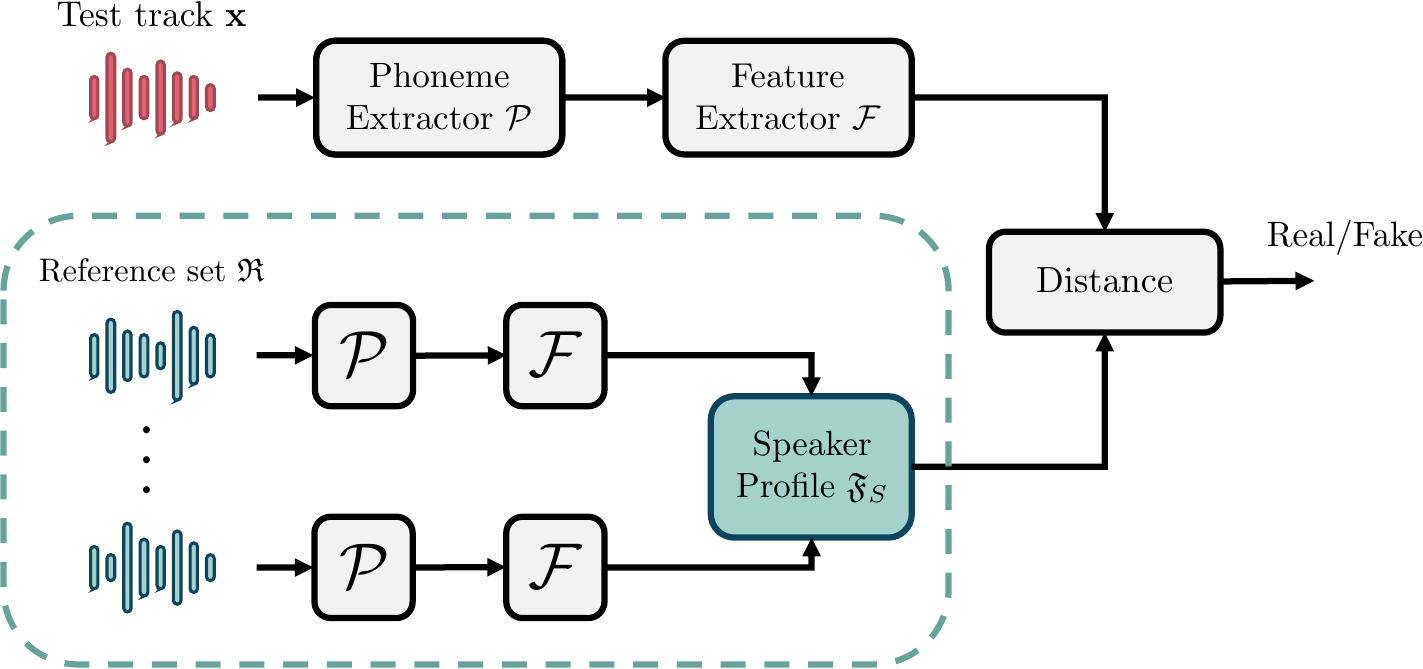}
    \caption{Pipeline of the proposed phoneme-level \gls{poi} speech deepfake detection method.}
    \label{fig:pipeline}
\end{figure}

In this paper, we propose a phoneme-based approach to \gls{poi} speech deepfake detection.
Our detection pipeline, illustrated in Figure~\ref{fig:pipeline}, is composed of the following steps:

\noindent \textbf{Phoneme Extraction}: Each speech signal $\mathbf{x}$ (including both reference and test tracks) is processed by a \textit{Phoneme Extractor} $\mathcal{P}$ to segment it into its constituent phonemes.
For each phoneme $a$ in the phonetic alphabet $\mathfrak{A}$, which includes all the \gls{ipa} phonemes, we define the set $\mathfrak{P}_\mathbf{x}^a = \{p^a_0, p^a_1, \dots, p^a_{N-1}\}$ that comprises all the realizations of the phoneme $a$ in $\mathbf{x}$. This segmentation enables fine-grained analysis of the speech signal.


\noindent \textbf{Feature Extraction}: Each phoneme instance $p^a_i$ is processed by a \textit{Feature Extractor} $\mathcal{F}$, generating a $d$-dimensional feature vector $\mathbf{f}^a_i \in \mathbb{R}^d$. This results in a set $\mathfrak{F}^a_\mathbf{x}$ which contains features corresponding to all phonemes in $\mathfrak{P}^a_\mathbf{x}$. Each embedding $\mathbf{f}^a_i$ encodes speaker-specific phoneme characteristics that are critical to distinguish between real and fake speech.
To represent the entire track $\mathbf{x}$ we aggregate all the sets $\mathfrak{F}^a_\mathbf{x}$ as in $\mathfrak{F}_\mathbf{x} = \left\{ \mathfrak{F}^a_\mathbf{x} \right\}_{a \in \mathfrak{A}}$.

\noindent \textbf{Speaker Profile Construction}: To build a target speaker profile, we repeat the phoneme and feature extraction steps across a set of reference tracks $\mathfrak{R}$ from the person of interest.
This process produces the speaker profile set $\mathfrak{F}_S$, which captures the phoneme-level acoustic patterns unique to the target speaker $S$. 

\noindent \textbf{Inference}: Given a test track $\mathbf{x}$, we compute its feature representation $\mathfrak{F}_\mathbf{x}$ as described above.
For each phoneme $a$, the feature vector $\mathbf{f}^a_i$ is compared against the corresponding phoneme representation in $\mathfrak{F}_S$ using a distance function, measuring how closely each test phoneme matches the reference speaker’s typical realization. 

\noindent \textbf{Classification}: Phoneme-level distances are aggregated to compute a global similarity score between the test signal and the speaker profile. This score quantifies the overall deviation of $\mathbf{x}$ from the expected voice pattern of speaker $S$.
This score is then used to classify the test sample as either authentic ($y=0$) or synthetic ($y=1$).

Our phoneme-level strategy offers several advantages over traditional speech deepfake detection methods. 
While conventional models provide only a binary classification, our framework can identify the specific phonemes and time intervals where synthetic artifacts occur.
This fine-grained analysis enhances interpretability, a critical aspect in multimedia forensics.
Additionally, by focusing only on the most relevant portions of the signal itself, i.e., phonemes, rather than analyzing the entire track, our method is inherently robust to post-processing operations that inject noise in the recording or apply lossy compression, as we will show in Section~\ref{subsec:robustness}.

\section{Experimental Setup}
\label{sec:setup}

In this section, we detail the experimental setup used in our analyses.
We begin by describing the speech processing models we used for phoneme extraction and feature computation.
Next, we outline the methodology for constructing speaker profiles and measuring distances between test tracks and reference profiles, along with the baseline we used to validate our findings.
Finally, we introduce the datasets used in our experiments.

\subsection{Speech Processing Models}
\label{subsec:speech_proc}

To process the input speech data, we employed two distinct pre-trained instances of Wav2Vec 2.0~\cite{baevski2020wav2vec}, a state-of-the-art self-supervised learning model speech representation.
The first model is the Phoneme Extractor $\mathcal{P}$ and consists of a fine-tuned version of Wav2Vec 2.0 on the LJSpeech Phonemes dataset~\cite{ljspeech_phonemes}, enabling it to predict phoneme sequences from raw audio waveforms.
The model processes the speech signal by dividing it into \SI{25}{\milli\second} frames with \SI{5}{\milli\second} overlap and assigns a phoneme to each frame where applicable.
The second model is the Feature Extractor $\mathcal{F}$ and consists of the \textit{base} version of Wav2Vec 2.0. It operates following the same frame-based processing pipeline as $\mathcal{P}$ but outputs a dense feature vector of dimension $d = 768$ for each frame, encoding rich acoustic information.

The two models operate sequentially: the input speech track $\mathbf{x}$ is first processed by $\mathcal{P}$ to extract phonemes and then passed through $\mathcal{F}$ to compute feature vectors. 
Since the two models share the same architecture and processing pipeline, their outputs are inherently time-aligned.
This allows us to directly associate each phoneme prediction from $\mathcal{P}$ with a corresponding feature vector from $\mathcal{F}$.

We retain only those feature vectors that are aligned with phoneme-labeled frames, discarding features corresponding to non-linguistic segments such as silence or unvoiced frames.
In cases where a phoneme spans multiple consecutive frames, the associated feature vectors are averaged to produce a single representative embedding for that phoneme.
Prior to model inference, all audio signals are normalized to have zero mean and unit variance. This standardization step reduces variability in amplitude and dynamic range across the tracks, enabling more robust phoneme prediction and feature extraction.

\subsection{Speaker profile and distance computation}

For each speaker $S$, we construct a profile $\mathfrak{F}_S$ in the form of a dictionary, where each key represents a phoneme $a$, and the associated value is the set of feature vectors $\mathfrak{F}^a_S$ for that phoneme across all the reference tracks, as in
\begin{equation}
        \mathfrak{F}_S = \left\{ a \mapsto \mathfrak{F}_S^a \mid a \in \mathfrak{A}_S \right\} ,
\end{equation}
where $\mathfrak{A}_S \subseteq \mathfrak{A}$ denotes the set of phonemes realized by the speaker $S$ in the reference tracks.

The number of reference utterances we use to construct each speaker profile is $I=100$.
This value is selected as a trade-off based on empirical validation, satisfying two main conditions.
The first one is phoneme coverage. With $I=100$ utterances per speaker, we ensure that most, if not all, phonemes in the speaker’s inventory are sufficiently represented in $\mathfrak{F}_S$, meaning that $\mathfrak{A}_S \approx \mathfrak{A}$.
Using fewer utterances may lead to data sparsity, where some phonemes are underrepresented or missing altogether.
The second condition is dataset balance, a practical consideration driven by experimental constraints. Limiting the number of reference tracks to $I=100$ allows the retention of a sufficient number of remaining utterances for test purposes across all speakers. Using a larger value for $I$ would disproportionately reduce the number of test samples available for some speakers, especially in datasets with limited total utterances per speaker.

During inference, the feature vector of each phoneme in a given test track $\mathbf{x}$ is compared to all corresponding phoneme entries in the speaker profile $\mathfrak{F}_S$.
Phoneme-level similarity is computed using the minimum element-wise cosine distance, following the approach of Pianese et al.~\cite{pianese2024training}.
This process is repeated for all phonemes in the test track, and the final distance between the test track $\mathbf{x}$ and the speaker profile $\mathfrak{F}_S$ is computed as the average of the individual phoneme distances.
Formally, for each phoneme occurrence $(p^a_i, \mathbf{f}^a_i)$, we compute the cosine distance to all vectors in the corresponding entry $\mathfrak{F}_S^{a}$ of the speaker profile, and retain the minimum distance, as in
\begin{equation}
    d_i = \min_{\mathbf{f} \in \mathfrak{F}_S^{a}} \left( 1 - \frac{\langle \mathbf{f}^a_i, \mathbf{f} \rangle}{\lVert \mathbf{f}^a_i \rVert \cdot \lVert \mathbf{f} \rVert} \right).
\end{equation}
Then, the overall distance between $\mathbf{x}$ and $\mathfrak{F}_S$ is obtained as
\begin{equation}
	D(\mathbf{x}, \mathfrak{F}_S) = \frac{1}{N} \sum_{n=0}^{N-1}{d_n},
\end{equation}
where $N$ is the number of phonemes contained in $\mathbf{x}$.
This final score $D(\mathbf{x}, \mathfrak{F}_S)$ is used as a decision metric for classifying $\mathbf{x}$ as either authentic or synthetic.

\subsection{Baseline}
\label{subsec:baselins}

To validate the effectiveness of the proposed pipeline, we compare it against a baseline method that follows the traditional \gls{poi} speech deepfake detection framework outlined in~\cite{pianese2024training}.
In this approach, each speech track $\mathbf{x}$ is represented by a single global feature vector $\mathbf{f}_\mathbf{x}$, computed as the average of the frame-level feature vectors generated by the feature extraction model $\mathcal{F}$, as in
\begin{equation}
	\mathbf{f}_\mathbf{x} = \frac{1}{T} \sum_{t=0}^{T-1}{\mathbf{f}_\mathbf{x}^t}, 
\end{equation}
where $T$ is the number of frames contained in $\mathbf{x}$.

Unlike our phoneme-level approach, which constructs a structured speaker profile as a dictionary indexed by phonemes, this baseline represents each speaker as a flat set of utterance-level feature vectors, each representing an entire reference signal, as in
\begin{equation}
	\mathfrak{F}_S = \left\{ \mathbf{f}_{\mathbf{x}_i} \in \mathbb{R}^d \mid i = 1, \dots, I \right\},
\end{equation}
where $I = 100$ is the number of reference utterances per speaker, consistent with our proposed method to ensure a fair comparison.

During inference, the similarity between a test track and a speaker profile is determined by computing the minimum cosine distance between the test track's feature vector $\mathbf{f}_\mathbf{x}$ and the feature vectors in $\mathfrak{F}_S$.
This distance serves as the final decision score, used to assess the authenticity of $\mathbf{x}$.

\subsection{Datasets}

We evaluate the proposed framework on four publicly available English-language speech deepfake datasets to provide a comprehensive assessment of its generalization capabilities across diverse conditions and synthesis techniques.
All audio data are sampled to a uniform sampling rate of $f_\text{s} = \SI{16}{\kilo\hertz}$. \vspace{3pt}

\noindent \textbf{ASVspoof 2019~\cite{todisco2019asvspoof}}.
Released for the homonymous challenge, this dataset was designed to develop effective \gls{asv} models.
We use its Logical Access-\textit{eval} partition, which includes real speech samples from VCTK~\cite{veaux2016superseded} and synthetic speech generated by \num{13} different speech synthesis models. 
The dataset features \num{67} speakers, enabling speaker-specific analyses. Detection performance is reported as the average across all speakers. \vspace{2pt}

\noindent \textbf{In-the-Wild~\cite{muller2022does}}. 
This dataset is designed to evaluate speech deepfake detectors in real-world conditions. It consists of approximately \num{38} hours of audio data (\num{17} hours fake, \num{21} hours real) featuring \num{54} celebrities and politicians.
The fake clips were created by segmenting publicly accessible video and audio files, while the real clips come from publicly available material featuring the same speakers. \vspace{2pt}

\noindent \textbf{Purdue speech dataset~\cite{bhagtani2024recent}}.
This corpus includes \num{25000} synthetic speech tracks generated by \num{5} diffusion model-based voice cloning methods: ProDiff, DiffGAN-TTS, ElevenLabs, UnitSpeech, and XTTS. Real speech data are sourced from LJSpeech~\cite{LJSpeech} and \num{10} speakers from LibriSpeech~\cite{panayotov2015librispeech}. \vspace{2pt}

\noindent \textbf{TIMIT-TTS~\cite{salvi2023timit}}.
This is a speech dataset that includes only fake audio samples, generated from \num{12} different \gls{tts} methods that reproduce the voice of Linda Johnson from LibriVox. It is created based on the VidTIMIT corpus~\cite{sanderson2002vidtimit} by generating a synthetic copy of its tracks using the considered synthetic speech generators. We consider the \textit{single speaker/clean} partition of this set, pairing it with real speech data from LJSpeech for evaluation.
\section{Results}
\label{sec:results}

In this section, we evaluate the performance of the proposed phoneme-based framework for speech deepfake detection.
The analysis aims to validate our hypothesis regarding the benefits of phoneme-level information for robustness and interpretability in the task at hand.

\subsection{Detection performance}

In our first experiment, we compare the speech deepfake detection performance of the proposed method against the considered baseline.
As described in Section~\ref{sec:setup}, both frameworks rely on the same feature encoder, i.e., the \textit{base} version of Wav2Vec 2.0, but differ in treating the input signal.
The proposed model incorporates a phoneme-level pre-processing step, while the baseline processes the entire signal without any linguistic structuring.

Table~\ref{tab:detection} shows the results of this analysis in terms of \gls{auc} and \gls{eer} across the four evaluated speech deepfake datasets.
The overall performance is comparable between the two methods.
The baseline achieves superior results on the ASVspoof 2019 dataset, while the proposed phoneme-level approach outperforms it on both the In-the-Wild and Purdue datasets.
For the TIMIT-TTS corpus, the two approaches yield nearly identical performance.

\begin{figure}
    \centering
    \includegraphics[width=0.95\columnwidth]{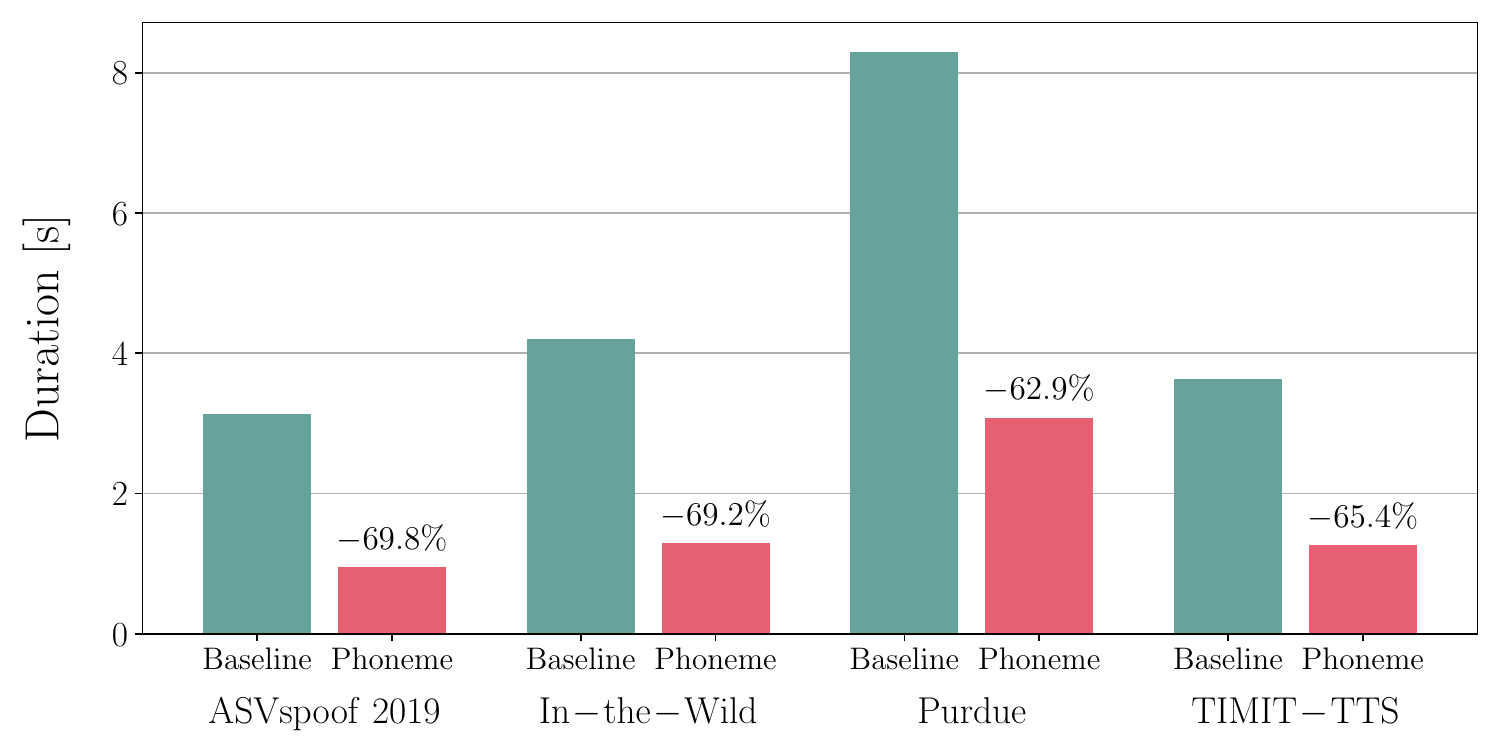}
    \caption{Comparison of average input signal duration (in seconds) processed by the baseline and proposed frameworks across the considered datasets.}
    \label{fig:duration}
    \vspace{-1em}
\end{figure}

Although the improvement in absolute detection performance might appear incremental, a deeper analysis reveals that it represents a significant finding.
Figure~\ref{fig:duration} shows that the proposed phoneme-based model processes, on average, over \num{65}\% less of the input signal during inference, while still achieving comparable or superior detection performance.
This efficiency stems from the core idea of our framework, which selectively analyzes only the portions of the signal associated with phonemes while discarding non-informative regions for the task at hand, such as silence or acoustically unclear content (see Section~\ref{subsec:speech_proc}).

This analysis also provides insight into the relatively poor performance of the proposed approach on the ASVspoof 2019 dataset with respect to the baseline (Table~\ref{tab:detection}).
In this specific corpus, the average amount of analyzed content per utterance is often shorter than one second, severely limiting the amount of information that can be used to perform the detection process.
This limitation, combined with known issues in ASVspoof 2019 regarding silences~\cite{muller2021speech} that may push the baseline performance, could be the reason behind the observed performance disparity.
On the other hand, our performance improves significantly on datasets containing longer test utterances, such as Purdue. Notably, this dataset contains the longest utterances among those evaluated and is also where our approach achieves the greatest performance improvement over the baseline.

These results indicate that the phoneme-based approach effectively isolates the most informative segments of the signal, capturing the critical aspects necessary for determining its authenticity.
The benefits are twofold: (i) a more compact and interpretable decision process and (ii) improved efficiency and potential scalability in practical deployment scenarios. 
We further investigate these advantages in the following experiments.

\begin{table}
\centering
\caption{Performance metrics (AUC and EER) for the considered methods, presented as percentage values (\%).}
\label{tab:detection}
\resizebox{\columnwidth}{!}{
\begin{tabular}{lcccccccc}
\toprule
         & \multicolumn{2}{c}{ASVspoof 2019}     & \multicolumn{2}{c}{In-the-Wild}       & \multicolumn{2}{c}{Purdue}       & \multicolumn{2}{c}{TIMIT-TTS} \\ \cmidrule(lr){2-3} \cmidrule(lr){4-5} \cmidrule(lr){6-7} \cmidrule(lr){8-9}
         & AUC $\uparrow$ & EER $\downarrow$ & AUC $\uparrow$ & EER $\downarrow$ & AUC $\uparrow$ & EER $\downarrow$ & AUC $\uparrow$ & EER $\downarrow$ \\ \midrule
        Baseline & \textbf{98.12}  & \textbf{6.84}   & 74.69          & 30.88          & 81.32          & 26.49          & 84.54 & \textbf{21.33} \\
        Ours (phoneme)           & 79.45           & 25.82  & \textbf{78.19} & \textbf{27.48} & \textbf{89.49} & \textbf{17.55} & \textbf{84.96} & 23.56          \\ \bottomrule
\end{tabular}}
\end{table}

\begin{table*}
\centering
\caption{Change in EER ($\Delta$EER, in \%) relative to clean test conditions, different post-processing operations are applied. Lower variation indicates greater robustness.}
\label{tab:robustness}
\resizebox{\textwidth}{!}{
\begin{tabular}{@{}lcccccccccccc@{}}
\toprule
     & \multicolumn{2}{c}{SNR = \SI{25}{\decibel}}  & \multicolumn{2}{c}{SNR = \SI{20}{\decibel}} & \multicolumn{2}{c}{SNR = \SI{15}{\decibel}} & \multicolumn{2}{c}{SNR = \SI{10}{\decibel}} & \multicolumn{2}{c}{MP3 compression} & \multicolumn{2}{c}{$\mu$-law quantization} \\ 
     \cmidrule(lr){2-3} \cmidrule(lr){4-5} \cmidrule(lr){6-7} \cmidrule(lr){8-9} \cmidrule(lr){10-11} \cmidrule(lr){12-13}
     & \multicolumn{1}{l}{Baseline} & \multicolumn{1}{l}{Ours (phoneme)} & \multicolumn{1}{l}{Baseline} & \multicolumn{1}{l}{Ours (phoneme)} & \multicolumn{1}{l}{Baseline} & \multicolumn{1}{l}{Ours (phoneme)} &
     \multicolumn{1}{l}{Baseline} & \multicolumn{1}{l}{Ours (phoneme)} &
     \multicolumn{1}{l}{Baseline} & \multicolumn{1}{l}{Ours (phoneme)} &
     \multicolumn{1}{l}{Baseline} & \multicolumn{1}{l}{Ours (phoneme)} \\ \midrule
ASVspoof 2019     & 13.69 & \textbf{0.20} & 14.66 & \textbf{0.40} & 15.46 & \textbf{0.68} & 12.97 & \textbf{5.06} & -0.08 & \textbf{-0.24}  & 0.08 & \textbf{-0.56} \\
In-the-Wild       & 5.24 & \textbf{-0.34} & 5.34 & \textbf{0.17} & 5.18 & \textbf{2.81} & \textbf{4.54} & 6.77 & 2.50 & \textbf{1.10} & 0.63 & \textbf{-0.08} \\
Purdue           & 7.95 & \textbf{0.00} & 11.59 & \textbf{1.32} & 10.60 & \textbf{3.97} & 13.91 & \textbf{5.24} & 7.42 & \textbf{6.74} & 4.95 & \textbf{3.64} \\
TIMIT-TTS         & 0.00 & \textbf{-1.66} & 0.67 & \textbf{-3.52} & 3.34   & \textbf{-4.80}  & 4.67 & \textbf{-2.74} & 0.91 & \textbf{-1.04} & 1.21 & \textbf{-0.12}            \\ \midrule
Average           & 6.72  & \textbf{-0.45} & 8.07 & \textbf{-0.41} & 8.65  & \textbf{0.67} & 9.02 & \textbf{3.58} & 2.69 & \textbf{1.64} & 1.72 & \textbf{0.72}               \\ \bottomrule
\end{tabular}}
\end{table*}

\subsection{Robustness to Post-Processing Perturbations}
\label{subsec:robustness}

We now evaluate the robustness of the proposed method to common post-processing perturbations, such as additive noise, lossy compression, and quantization artifacts, in comparison to the baseline system.
Our underlying hypothesis is that the phoneme-based framework, by focusing on the most informative and meaningful regions of the signal, is inherently more robust to degradations that predominantly affect less critical or non-phoneme regions.
In contrast, the baseline system, which processes the entire waveform indiscriminately, may extract features from segments that are more vulnerable to such perturbations, potentially degrading its performance under adverse conditions.

To test this hypothesis, we subject the test tracks to three types of distortions: additive Gaussian noise at various \gls{snr} levels ranging from \num{25} to \SI{10}{\decibel}, MP3 compression at a bitrate of 128 kbps, and 8-bit $\mu$-law quantization.
Processing is applied only to the test tracks, while the reference set remains clean.
This setup simulates realistic conditions for \gls{poi} methods, where clean reference recordings of the target speaker are assumed to be available, but test samples may be subject to noise or other distortions.

Table~\ref{tab:robustness} presents the results of this analysis in terms of \gls{eer} variations ($\Delta$\gls{eer}) relative to the clean condition.
The results confirm our hypothesis: the proposed method consistently outperforms the baseline in terms of robustness and exhibits lower sensitivity to noise.
Notably, the \gls{eer} increase of the phoneme-based method remains below \num{5}\% across almost all tested conditions, while the baseline exhibits an \gls{eer} increase exceeding \num{10}\% on both ASVspoof 2019 and Purdue datasets.

In some cases, the proposed framework even yields negative $\Delta$EER values, indicating not just robustness but actual performance improvements under distortion.
This may be attributed to the fact that after post-processing, only the clearest phonemes of the input signal are detected, while ambiguous or low-quality ones are filtered out, resulting in cleaner feature representations.
Also, we observe increasing performance degradation as the \gls{snr} decreases, suggesting that noise injection obscures information useful for distinguishing between real and fake samples.
On the other hand, both MP3 compression and $\mu$-law quantization introduce relatively minor performance drops. Nonetheless, the proposed method consistently outperforms the baseline under these conditions as well.

These findings suggest that phoneme-level modeling not only improves efficiency but also gives increased robustness against common distortions encountered in real-world scenarios.

\subsection{Interpretability analysis}

The final aspect we want to assess in the proposed method is its interpretability.
Since our framework performs a fine-grained analysis of the input signal and operates at the phoneme level, it is capable of identifying the specific phonemes that deviate from the speaker’s reference profile. This allows for temporal interpretability, as it highlights precisely when and where in the utterance anomalies occur.
These insights are particularly valuable in several forensic scenarios, such as courtrooms and legal proceedings, where explainability is essential for justifying detection outcomes. 
Understanding why a speech segment is classified as fake (e.g., because a specific phoneme is articulated differently from how the speaker typically pronounces it) provides highly actionable forensic evidence.

To investigate this interpretability aspect, we categorized all phonemes into seven groups based on the standard \gls{ipa} classification and evaluated the detection performance using only one category at a time.
The goal of this experiment is to determine which phoneme classes are most challenging for deepfake generators to replicate, and therefore, which contribute most to effective detection.
Table~\ref{tab:phonemes} summarizes the phoneme categories that we considered.

Table~\ref{tab:xai_values} presents the results of this analysis.
Vowels and plosives emerge as the most discriminative phoneme categories, consistent with findings from~\cite{dhamyal2021using}, while affricates appear to be the least informative.
Also, no single phoneme category achieves better performance than the combined set of all phonemes, indicating the benefit of comprehensive phonetic coverage.
An important observation is that the occurrence of phonemes in the analyzed data significantly impacts their utility. For instance, vowels occur far more frequently than diphthongs or approximants, leading to more robust speaker profiles and more reliable classification.
In future work, we plan to explore strategies to effectively utilize all phoneme categories, regardless of their frequency in the analyzed speech.

\begin{table}
\centering
\resizebox{.9\columnwidth}{!}{
\begin{tabular}{@{}ll@{}}
\toprule
\textbf{Category}      & \textbf{Phonemes} \\
\midrule
\textit{Vowels}        & /\textipa{i}/, /\textipa{æ}/, /\textipa{A}/, /\textipa{I}/, /\textipa{V}/, /\textipa{u}/, /\textipa{O}/, /\textipa{@}/, /\textipa{3r}/, /\textipa{E}/, /\textipa{U}/ \\
\textit{Diphthongs}    & /\textipa{aI}/, /\textipa{oU}/, /\textipa{OI}/, /\textipa{aU}/, /\textipa{eI}/ \\
\textit{Plosives}      & /\textipa{k}/, /\textipa{p}/, /\textipa{t}/, /\textipa{b}/, /\textipa{d}/, /\textipa{g}/ \\
\textit{Fricatives}    & /\textipa{S}/, /\textipa{s}/, /\textipa{z}/, /\textipa{T}/, /\textipa{D}/, /\textipa{f}/, /\textipa{v}/, /\textipa{Z}/, /\textipa{h}/ \\
\textit{Affricates}    & /\textipa{tS}/, /\textipa{dZ}/ \\
\textit{Approximants}  & /\textipa{l}/, /\textipa{j}/, /\textipa{w}/, /\textipa{r}/ \\
\textit{Nasals}        & /\textipa{m}/, /\textipa{n}/, /\textipa{N}/ \\
\bottomrule
\end{tabular}}
\caption{Phoneme categories considered in the analysis}
\label{tab:phonemes}
\end{table}

\begin{table}
\centering
\caption{Detection performance per phoneme category in terms of EER ($\downarrow$). Best discriminative categories are highlighted in bold green, worst in italic red.}
\label{tab:xai_values}
\resizebox{\columnwidth}{!}{
\begin{tabular}{@{}lccccc@{}}
\toprule
             & ASVspoof 2019                         & In-the-Wild                           & Purdue                                & TIMIT-TTS                             & Average \\ \midrule 
All phonemes & 25.82                                 & 27.48                                 & 17.55                                 & 23.56                                 & 23.60                                 \\ \midrule
Vowels       & 32.85                  & {\color{OliveGreen} \textbf{29.54}}                        & {\color{OliveGreen} \textbf{28.59}}  & 30.22 & {\color{OliveGreen} \textbf{30.30} }                       \\
Diphthongs   & 38.36                                 & 36.99                                 & 39.42                                 & 34.25                                 & 37.26                                 \\
Plosives     & {\color{OliveGreen} \textbf{32.29}} & 31.60                        & 32.74 & {\color{OliveGreen}  \textbf{25.80}} & 30.61                        \\
Fricatives   & 33.47                                 & 34.19                                 & 36.61                                 & 40.51                                 & 36.20                                 \\
Affricates   & {\color{BrickRed} \textit{40.57}} & {\color{BrickRed} \textit{41.53}} & {\color{BrickRed} \textit{46.71}} & {\color{BrickRed} \textit{43.02}} & {\color{BrickRed} \textit{42.96}} \\
Approximants & 36.05                                 & 32.12                                 & 33.67                                 & 31.48                                 & 33.33                                 \\
Nasals       & 36.05                                 & 35.78                                 & 42.97                                 & 33.53                                 & 37.08 \\ \bottomrule
\end{tabular}}
\end{table}

\subsection{Case Study: Interpretability in Action}

To further illustrate the advantages of the proposed method in terms of interpretability, we conduct a focused case study involving two well-known speakers from the In-the-Wild dataset: Donald Trump and Queen Elizabeth II.
For each speaker, we analyze both a real and a synthetic version of the same utterance, applying our phoneme-level framework to compute distances between the test signals and the corresponding speaker-specific profiles.
The authentic recordings are sourced from the In-the-Wild dataset, as well as the tracks used to construct the speaker profiles.
On the other hand, the synthetic counterparts are generated using the ElevenLabs voice cloning tool to ensure identical linguistic content across the real and fake samples.

\begin{figure}
    \centering
    \begin{subfigure}[b]{\columnwidth}
        \centering
        \includegraphics[width=\textwidth]{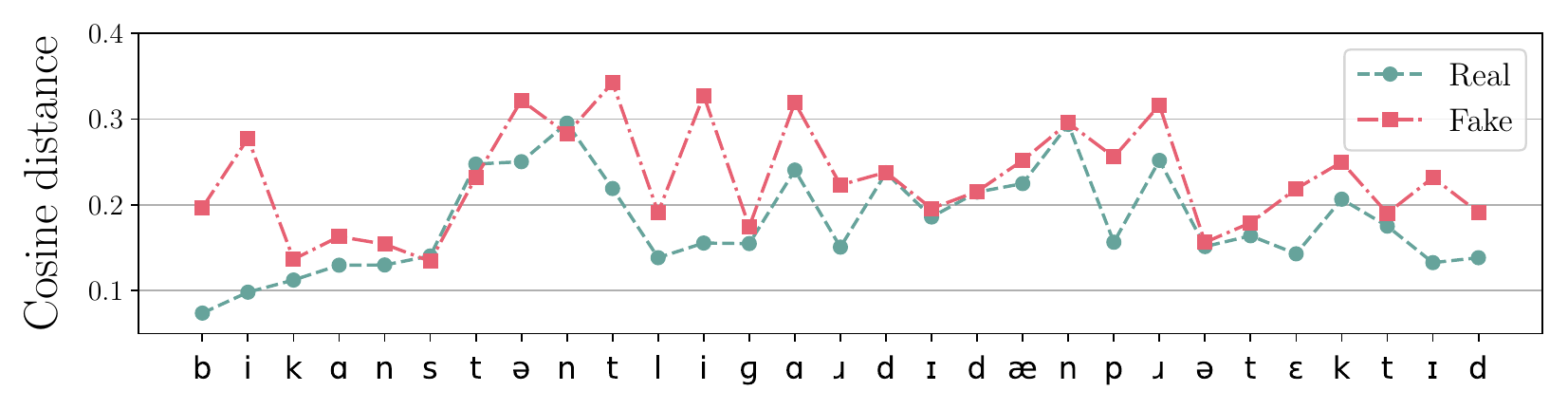}
        \caption{Donald Trump}
        \label{fig:trump}
    \end{subfigure}

    \vspace{0.5em} 

    \begin{subfigure}[b]{\columnwidth}
        \centering
        \includegraphics[width=\textwidth]{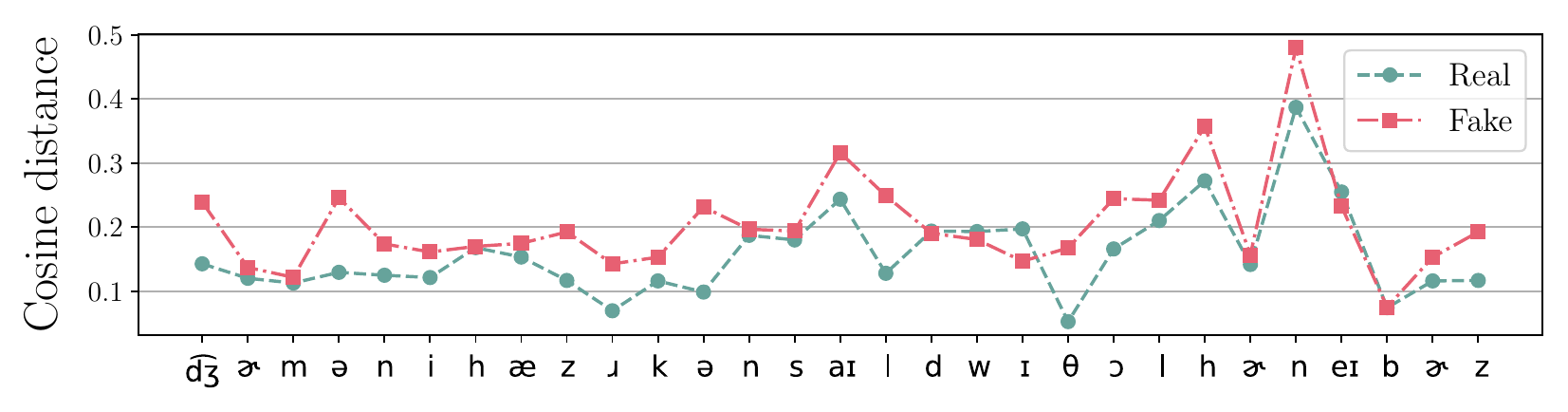}
        \caption{Queen Elizabeth II}
        \label{fig:queen}
    \end{subfigure}

    \caption{Case study: Phoneme-distance comparison between real and fake speech signals for two speakers.}
    \label{fig:case_study}
\end{figure}

The sentence analyzed for Donald Trump is \textit{``Be constantly guarded and protected''} (/\textipa{bi kAnst@ntli gArdId ænd pr@tEktId}/), and for Queen Elizabeth II it is \textit{``Germany has reconciled with all her neighbors''} (/\textipa{dZ3rm@ni hæz rk@nsaIld wIT Ol h3r neIb3rz}/).
Figure~\ref{fig:case_study} shows the phoneme-wise distance scores computed by our method for both the real and fake utterances, highlighting its capacity for detailed, interpretable analysis.

In general, the synthetic speech samples exhibit higher phoneme-level deviations from the reference profiles compared to authentic speech.
Additionally, the method identifies which specific phonemes contribute most to these deviations.
In the case of Donald Trump, the phonemes /\textipa{i}/ and /\textipa{A}/ show the largest mismatches, while for Queen Elizabeth II, the highest deviation is observed in the phoneme /\textipa{@}/.

This case study highlights the practical forensic utility of our approach. By localizing deviations at the phoneme level, the method may not only enable more precise deepfake detection but also provide interpretable evidence that can support human verification or judicial review.

\subsection{Discussion and Limitations}

The proposed phoneme-level \gls{poi}-based detection framework offers a novel perspective on speech deepfake detection by combining speaker-specific modeling with fine-grained phonetic analysis and produces promising results.

However, the method also presents some limitations that merit further investigation.
First, the framework requires access to a sufficiently large set of reference recordings from the target speaker to build a reliable phoneme-level profile. Ideally, this profile should include consistent coverage of all phoneme classes to ensure robustness during inference.

Second, the current framework depends on accurate phoneme segmentation, which is achieved through an external \gls{asr} system. Errors in alignment can propagate through the pipeline and negatively impact detection performance. As reliance on external components introduces an additional source of uncertainty and complexity, future work should explore ways to reduce or eliminate this dependency, such as through end-to-end alignment-free alternatives.

Related to this, the computational efficiency of the current pipeline is another area for improvement. At present, the method relies on the sequential inference of two distinct models, one for phoneme alignment and one for detection.
While this design enables modularity and interpretability, it is not optimal in terms of runtime performance.

Finally, our evaluation was limited to datasets containing only English speech.
The extension of our proposed method to other languages, especially those with different phoneme alphabets, may be nontrivial and is a subject of future research.

\section{Conclusions}
\label{sec:conclusion}

In this paper, we introduced a novel phoneme-level approach to \gls{poi}-based speech deepfake detection.
Our method decomposes speech signals into phonemes and constructs a speaker-specific phoneme profile using only real reference recordings.
During inference, each phoneme in a test utterance is independently compared to this profile to determine its authenticity, enabling a fine-grained and interpretable analysis.

This phoneme-centric design offers multiple advantages over traditional \gls{poi} approaches.
First, it enhances interpretability by revealing which specific phonetic elements deviate from the speaker’s speech patterns.
Second, it improves robustness, as the method focuses on the most relevant segments of the signal and is less prone to performance degradation in the case of post-processing.

Our findings highlight the potential of phoneme-level analysis as a promising direction for future research in speech deepfake detection. Future work could explore integrating phonetic analysis into end-to-end pipelines, further improving the robustness and interpretability of detection systems, as well as extending the framework to multilingual scenarios.
Additionally, the development of more sophisticated phoneme-level speaker profiles, potentially incorporating speaker-specific traits such as accent, intonation, or prosodic patterns, could further increase the accuracy and reliability of speaker-aware deepfake detection systems.

\section*{Acknowledgment}
This work was supported by the FOSTERER project, funded by the Italian Ministry of Education, University, and Research within the PRIN 2022 program. This work was partially supported by the European Union - Next Generation EU under the Italian National Recovery and Resilience Plan (NRRP), Mission 4, Component 2, Investment 1.3, CUP D43C22003080001, partnership on ``Telecommunications of the Future’’ (PE00000001 - program ``RESTART’’) and ``SEcurity and RIghts in the CyberSpace’’ (PE00000014 - program ``FF4ALL-SERICS’’).

{
    \small
    \bibliographystyle{ieeenat_fullname}
    \bibliography{main}

\begin{thebibliography}{48}
\providecommand{\natexlab}[1]{#1}
\providecommand{\url}[1]{\texttt{#1}}
\expandafter\ifx\csname urlstyle\endcsname\relax
  \providecommand{\doi}[1]{doi: #1}\else
  \providecommand{\doi}{doi: \begingroup \urlstyle{rm}\Url}\fi

\bibitem[Agarwal et~al.(2019)Agarwal, Farid, Gu, He, Nagano, and Li]{agarwal2019protecting}
Shruti Agarwal, Hany Farid, Yuming Gu, Mingming He, Koki Nagano, and Hao Li.
\newblock Protecting world leaders against deep fakes.
\newblock In \emph{IEEE/CVF Conference on Computer Vision and Pattern Recognition Workshop (CVPRW)}, 2019.

\bibitem[Agarwal et~al.(2020)Agarwal, Farid, Fried, and Agrawala]{agarwal2020detecting}
Shruti Agarwal, Hany Farid, Ohad Fried, and Maneesh Agrawala.
\newblock Detecting deep-fake videos from phoneme-viseme mismatches.
\newblock In \emph{IEEE/CVF Conference on Computer Vision and Pattern Recognition Workshops (CVPRW)}, 2020.

\bibitem[Amerini et~al.(2025)Amerini, Barni, Battiato, Bestagini, Boato, Bruni, Caldelli, De~Natale, De~Nicola, Guarnera, et~al.]{amerini2025deepfake}
Irene Amerini, Mauro Barni, Sebastiano Battiato, Paolo Bestagini, Giulia Boato, Vittoria Bruni, Roberto Caldelli, Francesco De~Natale, Rocco De~Nicola, Luca Guarnera, et~al.
\newblock Deepfake media forensics: Status and future challenges.
\newblock \emph{Journal of Imaging}, 11\penalty0 (3):\penalty0 73, 2025.

\bibitem[Attorresi et~al.(2022)Attorresi, Salvi, Borrelli, Bestagini, and Tubaro]{attorresi2022prosody}
Luigi Attorresi, Davide Salvi, Clara Borrelli, Paolo Bestagini, and Stefano Tubaro.
\newblock {Combining Automatic Speaker Verification and Prosody Analysis for Synthetic Speech Detection}.
\newblock In \emph{International Conference on Pattern Recognition (ICPR)}, 2022.

\bibitem[Baevski et~al.(2020)Baevski, Zhou, Mohamed, and Auli]{baevski2020wav2vec}
Alexei Baevski, Yuhao Zhou, Abdelrahman Mohamed, and Michael Auli.
\newblock {wav2vec 2.0: A framework for self-supervised learning of speech representations}.
\newblock \emph{Advances in Neural Information Processing Systems (NeurIPS)}, 2020.

\bibitem[Bartusiak et~al.(2023)Bartusiak, Bhagtani, Yadav, and Delp]{bartusiak2023transformer}
Emily~R Bartusiak, Kratika Bhagtani, Amit Kumar~Singh Yadav, and Edward~J Delp.
\newblock Transformer ensemble for synthesized speech detection.
\newblock In \emph{Asilomar Conference on Signals, Systems, and Computers}. IEEE, 2023.

\bibitem[Bhagtani et~al.(2024)Bhagtani, Yadav, Bestagini, and Delp]{bhagtani2024recent}
Kratika Bhagtani, Amit Kumar~Singh Yadav, Paolo Bestagini, and Edward~J Delp.
\newblock {Are Recent Deepfake Speech Generators Detectable?}
\newblock In \emph{ACM Workshop on Information Hiding and Multimedia Security}, 2024.

\bibitem[Bookbot(2023)]{ljspeech_phonemes}
Bookbot.
\newblock Ljspeech phonemes dataset.
\newblock \url{https://huggingface.co/datasets/bookbot/ljspeech_phonemes}, 2023.
\newblock Accessed: 2025-02-14.

\bibitem[Chan et~al.(2016)Chan, Jaitly, Le, and Vinyals]{chan2016listen}
William Chan, Navdeep Jaitly, Quoc Le, and Oriol Vinyals.
\newblock Listen, attend and spell: A neural network for large vocabulary conversational speech recognition.
\newblock In \emph{IEEE International Conference on Acoustics, Speech and Signal Processing (ICASSP)}, 2016.

\bibitem[Coletta et~al.(2025)Coletta, Salvi, Negroni, Leonzio, and Bestagini]{coletta2025anomaly}
Emma Coletta, Davide Salvi, Viola Negroni, Daniele~Ugo Leonzio, and Paolo Bestagini.
\newblock Anomaly detection and localization for speech deepfakes via feature pyramid matching.
\newblock \emph{arXiv preprint arXiv:2503.18032}, 2025.

\bibitem[Cozzolino et~al.(2021)Cozzolino, R{\"o}ssler, Thies, Nie{\ss}ner, and Verdoliva]{cozzolino2021id}
Davide Cozzolino, Andreas R{\"o}ssler, Justus Thies, Matthias Nie{\ss}ner, and Luisa Verdoliva.
\newblock {Id-reveal: Identity-aware deepfake video detection}.
\newblock In \emph{IEEE/CVF Conference on Computer Vision and Pattern Recognition (CVPR)}, 2021.

\bibitem[Cozzolino et~al.(2023)Cozzolino, Pianese, Nie{\ss}ner, and Verdoliva]{cozzolino2023audio}
Davide Cozzolino, Alessandro Pianese, Matthias Nie{\ss}ner, and Luisa Verdoliva.
\newblock Audio-visual person-of-interest deepfake detection.
\newblock In \emph{IEEE/CVF Conference on Computer Vision and Pattern Recognition (CVPR)}, 2023.

\bibitem[Cuccovillo et~al.(2022)Cuccovillo, Papastergiopoulos, Vafeiadis, Yaroshchuk, Aichroth, Votis, and Tzovaras]{cuccovillo2022open}
Luca Cuccovillo, Christoforos Papastergiopoulos, Anastasios Vafeiadis, Artem Yaroshchuk, Patrick Aichroth, Konstantinos Votis, and Dimitrios Tzovaras.
\newblock Open challenges in synthetic speech detection.
\newblock In \emph{IEEE International Workshop on Information Forensics and Security (WIFS)}, 2022.

\bibitem[Cuccovillo et~al.(2024)Cuccovillo, Gerhardt, and Aichroth]{cuccovillo2024audio}
Luca Cuccovillo, Milica Gerhardt, and Patrick Aichroth.
\newblock Audio transformer for synthetic speech detection via multi-formant analysis.
\newblock In \emph{IEEE/CVF Conference on Computer Vision and Pattern Recognition Workshops (CVPRW)}, 2024.

\bibitem[Dhamyal et~al.(2021)Dhamyal, Ali, Qazi, and Raza]{dhamyal2021using}
Hira Dhamyal, Ayesha Ali, Ihsan~Ayyub Qazi, and Agha~Ali Raza.
\newblock {Using self attention DNNs to discover phonemic features for audio deep fake detection}.
\newblock In \emph{IEEE Automatic Speech Recognition and Understanding Workshop (ASRU)}, 2021.

\bibitem[Ge et~al.(2023)Ge, Tak, Todisco, and Evans]{ge2023can}
Wanying Ge, Hemlata Tak, Massimiliano Todisco, and Nicholas Evans.
\newblock Can spoofing countermeasure and speaker verification systems be jointly optimised?
\newblock In \emph{IEEE International Conference on Acoustics, Speech and Signal Processing (ICASSP)}, 2023.

\bibitem[Golda et~al.(2024)Golda, Mekonen, Pandey, Singh, Hassija, Chamola, and Sikdar]{golda2024privacy}
Abenezer Golda, Kidus Mekonen, Amit Pandey, Anushka Singh, Vikas Hassija, Vinay Chamola, and Biplab Sikdar.
\newblock Privacy and security concerns in generative ai: A comprehensive survey.
\newblock \emph{IEEE Access}, 2024.

\bibitem[Ito and Johnson(2017)]{LJSpeech}
Keith Ito and Linda Johnson.
\newblock {The LJ Speech Dataset}.
\newblock \url{https://keithito.com/LJ-Speech-Dataset/}, 2017.

\bibitem[Jung et~al.(2022)Jung, Heo, Tak, Shim, Chung, Lee, Yu, and Evans]{jung2022aasist}
Jee-weon Jung, Hee-Soo Heo, Hemlata Tak, Hye-jin Shim, Joon~Son Chung, Bong-Jin Lee, Ha-Jin Yu, and Nicholas Evans.
\newblock Aasist: Audio anti-spoofing using integrated spectro-temporal graph attention networks.
\newblock In \emph{IEEE International Conference on Acoustics, Speech and Signal Processing (ICASSP)}, 2022.

\bibitem[Li et~al.(2025)Li, Ahmadiadli, and Zhang]{li2025survey}
Menglu Li, Yasaman Ahmadiadli, and Xiao-Ping Zhang.
\newblock A survey on speech deepfake detection.
\newblock \emph{ACM Computing Surveys}, 2025.

\bibitem[Ma et~al.(2018)Ma, Sethu, Ambikairajah, and Lee]{ma2018speaker}
Jianbo Ma, Vidhyasaharan Sethu, Eliathamby Ambikairajah, and Kong~Aik Lee.
\newblock Speaker-phonetic vector estimation for short duration speaker verification.
\newblock In \emph{IEEE International Conference on Acoustics, Speech and Signal Processing (ICASSP)}, 2018.

\bibitem[Masood et~al.(2023)Masood, Nawaz, Malik, Javed, Irtaza, and Malik]{masood2023deepfakes}
Momina Masood, Mariam Nawaz, Khalid~Mahmood Malik, Ali Javed, Aun Irtaza, and Hafiz Malik.
\newblock Deepfakes generation and detection: State-of-the-art, open challenges, countermeasures, and way forward.
\newblock \emph{Applied intelligence}, 53\penalty0 (4):\penalty0 3974--4026, 2023.

\bibitem[M{\"u}ller et~al.(2021)M{\"u}ller, Dieckmann, Czempin, Canals, B{\"o}ttinger, and Williams]{muller2021speech}
Nicolas~M M{\"u}ller, Franziska Dieckmann, Pavel Czempin, Roman Canals, Konstantin B{\"o}ttinger, and Jennifer Williams.
\newblock {Speech is silver, silence is golden: What do ASVspoof-trained models really learn?}
\newblock In \emph{Interspeech}, 2021.

\bibitem[M{\"u}ller et~al.(2022)M{\"u}ller, Czempin, Dieckmann, Froghyar, and B{\"o}ttinger]{muller2022does}
Nicolas~M M{\"u}ller, Pavel Czempin, Franziska Dieckmann, Adam Froghyar, and Konstantin B{\"o}ttinger.
\newblock Does audio deepfake detection generalize?
\newblock In \emph{Interspeech}, 2022.

\bibitem[Nagarsheth et~al.(2017)Nagarsheth, Khoury, Patil, and Garland]{nagarsheth2017replay}
Parav Nagarsheth, Elie Khoury, Kailash Patil, and Matt Garland.
\newblock Replay attack detection using dnn for channel discrimination.
\newblock In \emph{Interspeech}, 2017.

\bibitem[Negroni et~al.(2025)Negroni, Salvi, Mezza, Bestagini, and Tubaro]{negroni2025leveraging}
Viola Negroni, Davide Salvi, Alessandro~Ilic Mezza, Paolo Bestagini, and Stefano Tubaro.
\newblock Leveraging mixture of experts for improved speech deepfake detection.
\newblock In \emph{IEEE International Conference on Acoustics, Speech and Signal Processing (ICASSP)}, 2025.

\bibitem[Palaskar and Metze(2018)]{palaskar2018acoustic}
Shruti Palaskar and Florian Metze.
\newblock Acoustic-to-word recognition with sequence-to-sequence models.
\newblock In \emph{IEEE Spoken Language Technology Workshop (SLT)}, 2018.

\bibitem[Panayotov et~al.(2015)Panayotov, Chen, Povey, and Khudanpur]{panayotov2015librispeech}
Vassil Panayotov, Guoguo Chen, Daniel Povey, and Sanjeev Khudanpur.
\newblock Librispeech: an {ASR} corpus based on public domain audio books.
\newblock In \emph{IEEE International Conference on Acoustics, Speech and Signal Processing (ICASSP)}, 2015.

\bibitem[Pianese et~al.(2022)Pianese, Cozzolino, Poggi, and Verdoliva]{pianese2022deepfake}
Alessandro Pianese, Davide Cozzolino, Giovanni Poggi, and Luisa Verdoliva.
\newblock Deepfake audio detection by speaker verification.
\newblock In \emph{IEEE International Workshop on Information Forensics and Security (WIFS)}, 2022.

\bibitem[Pianese et~al.(2024)Pianese, Cozzolino, Poggi, and Verdoliva]{pianese2024training}
Alessandro Pianese, Davide Cozzolino, Giovanni Poggi, and Luisa Verdoliva.
\newblock Training-free deepfake voice recognition by leveraging large-scale pre-trained models.
\newblock In \emph{ACM Workshop on Information Hiding and Multimedia Security}, 2024.

\bibitem[Rana et~al.(2024)Rana, Nobi, Murali, and Sung]{rana2022deepfake}
Md~Shohel Rana, Mohammad~Nur Nobi, Beddhu Murali, and Andrew~H Sung.
\newblock Deepfake detection: A systematic literature review.
\newblock \emph{IEEE Access}, 2024.

\bibitem[Salvi et~al.(2023)Salvi, Hosler, Bestagini, Stamm, and Tubaro]{salvi2023timit}
Davide Salvi, Brian Hosler, Paolo Bestagini, Matthew~C Stamm, and Stefano Tubaro.
\newblock {TIMIT-TTS: a Text-to-Speech Dataset for Multimodal Synthetic Media Detection}.
\newblock \emph{IEEE Access}, 2023.

\bibitem[Sanderson(2002)]{sanderson2002vidtimit}
Conrad Sanderson.
\newblock {The VidTIMIT database}.
\newblock Technical report, IDIAP, 2002.

\bibitem[Schwarz et~al.(2006)Schwarz, Matejka, and Cernocky]{schwarz2006hierarchical}
Petr Schwarz, Pavel Matejka, and Jan Cernocky.
\newblock Hierarchical structures of neural networks for phoneme recognition.
\newblock In \emph{IEEE International Conference on Acoustics Speech and Signal Processing (ICASSP)}, 2006.

\bibitem[Sethu et~al.(2008)Sethu, Ambikairajah, and Epps]{sethu2008phonetic}
Vidhyasaharan Sethu, Eliathamby Ambikairajah, and Julien Epps.
\newblock Phonetic and speaker variations in automatic emotion classification.
\newblock In \emph{Interspeech}, 2008.

\bibitem[Suthokumar et~al.(2019)Suthokumar, Sriskandaraja, Sethu, Wijenayake, and Ambikairajah]{suthokumar2019phoneme}
Gajan Suthokumar, Kaavya Sriskandaraja, Vidhyasaharan Sethu, Chamith Wijenayake, and Eliathamby Ambikairajah.
\newblock Phoneme specific modelling and scoring techniques for anti spoofing system.
\newblock In \emph{IEEE International Conference on Acoustics, Speech and Signal Processing (ICASSP)}, 2019.

\bibitem[Tak et~al.(2022)Tak, Todisco, Wang, Jung, Yamagishi, and Evans]{tak2022automatic}
Hemlata Tak, Massimiliano Todisco, Xin Wang, Jee-weon Jung, Junichi Yamagishi, and Nicholas Evans.
\newblock Automatic speaker verification spoofing and deepfake detection using wav2vec 2.0 and data augmentation.
\newblock In \emph{Interspeech}, 2022.

\bibitem[Todisco et~al.(2019)Todisco, Wang, Vestman, Sahidullah, Delgado, Nautsch, Yamagishi, Evans, Kinnunen, and Lee]{todisco2019asvspoof}
Massimiliano Todisco, Xin Wang, Ville Vestman, Md Sahidullah, H{\'e}ctor Delgado, Andreas Nautsch, Junichi Yamagishi, Nicholas Evans, Tomi Kinnunen, and Kong~Aik Lee.
\newblock {ASVspoof 2019: Future horizons in spoofed and fake audio detection}.
\newblock In \emph{Interspeech}, 2019.

\bibitem[Veaux et~al.(2016)Veaux, Yamagishi, MacDonald, et~al.]{veaux2016superseded}
Christophe Veaux, Junichi Yamagishi, Kirsten MacDonald, et~al.
\newblock {Superseded-CSTR VCTK Corpus: English Multi-Speaker Corpus for CSTR Voice Cloning Toolkit}.
\newblock \emph{University of Edinburgh. The Centre for Speech Technology Research (CSTR)}, 2016.

\bibitem[Verdoliva(2020)]{verdoliva2020media}
Luisa Verdoliva.
\newblock {Media forensics and deepfakes: an overview}.
\newblock \emph{IEEE Journal of Selected Topics in Signal Processing}, 14\penalty0 (5):\penalty0 910--932, 2020.

\bibitem[Wang et~al.(2025)Wang, Delgado, Tak, Jung, Shim, Todisco, Kukanov, Liu, Sahidullah, Kinnunen, et~al.]{wang2025asvspoof}
Xin Wang, H{\'e}ctor Delgado, Hemlata Tak, Jee-weon Jung, Hye-jin Shim, Massimiliano Todisco, Ivan Kukanov, Xuechen Liu, Md Sahidullah, Tomi Kinnunen, et~al.
\newblock Asvspoof 5: Design, collection and validation of resources for spoofing, deepfake, and adversarial attack detection using crowdsourced speech.
\newblock \emph{Computer Speech \& Language}, page 101825, 2025.

\bibitem[Wang et~al.(2024)Wang, Fu, Wen, Tao, Wang, Xie, Qi, Shi, Lu, Liu, et~al.]{wang2024mixture}
Zhiyong Wang, Ruibo Fu, Zhengqi Wen, Jianhua Tao, Xiaopeng Wang, Yuankun Xie, Xin Qi, Shuchen Shi, Yi Lu, Yukun Liu, et~al.
\newblock Mixture of experts fusion for fake audio detection using frozen wav2vec 2.0.
\newblock \emph{arXiv preprint arXiv:2409.11909}, 2024.

\bibitem[weon Jung et~al.(2024)weon Jung, Wang, Evans, Watanabe, jin Shim, Tak, Arora, Yamagishi, and Chung]{jung24d_interspeech}
Jee weon Jung, Xin Wang, Nicholas Evans, Shinji Watanabe, Hye jin Shim, Hemlata Tak, Siddhant Arora, Junichi Yamagishi, and Joon~Son Chung.
\newblock {To what extent can ASV systems naturally defend against spoofing attacks?}
\newblock In \emph{Interspeech}, 2024.

\bibitem[Yang et~al.(2024)Yang, Qin, Zhou, Wang, Guo, Han, and Wang]{yang2024robust}
Yujie Yang, Haochen Qin, Hang Zhou, Chengcheng Wang, Tianyu Guo, Kai Han, and Yunhe Wang.
\newblock A robust audio deepfake detection system via multi-view feature.
\newblock In \emph{IEEE International Conference on Acoustics, Speech and Signal Processing (ICASSP)}, 2024.

\bibitem[Zaman et~al.(2024)Zaman, Samiul, Sah, Direkoglu, Okada, and Unoki]{zaman2024hybrid}
Khalid Zaman, Islam~JAM Samiul, Melike Sah, Cem Direkoglu, Shogo Okada, and Masashi Unoki.
\newblock Hybrid transformer architectures with diverse audio features for deepfake speech classification.
\newblock \emph{IEEE Access}, 2024.

\bibitem[Zhang et~al.(2025)Zhang, Hua, Lan, Zhang, and Guo]{zhang2025phoneme}
Kuiyuan Zhang, Zhongyun Hua, Rushi Lan, Yushu Zhang, and Yifang Guo.
\newblock Phoneme-level feature discrepancies: A key to detecting sophisticated speech deepfakes.
\newblock In \emph{AAAI Conference on Artificial Intelligence}, 2025.

\bibitem[Zhang et~al.(2024)Zhang, Wen, and Hu]{zhang2024audio}
Qishan Zhang, Shuangbing Wen, and Tao Hu.
\newblock Audio deepfake detection with self-supervised xls-r and sls classifier.
\newblock In \emph{ACM International Conference on Multimedia}, 2024.

\bibitem[Zhang et~al.(2021)Zhang, Jiang, and Duan]{zhang2021one}
You Zhang, Fei Jiang, and Zhiyao Duan.
\newblock One-class learning towards synthetic voice spoofing detection.
\newblock \emph{IEEE Signal Processing Letters}, 28:\penalty0 937--941, 2021.

\end{thebibliography}
}

\end{document}